\documentclass[a4paper]{article}
\usepackage[utf8]{inputenc}
\usepackage{amsmath,amssymb}
\usepackage{bbold}
\usepackage{xcolor}
\usepackage{multirow}

\def\z{{\mathbf z}}
\def\o{{\mathbf o}}

\def\h{{\mathbf h}}

\newcommand{\tb}[1]{\textbf{#1}}
\newcommand{\norm}[1]{\left\lVert#1\right\rVert}
\usepackage{INTERSPEECH2022}

\title{Training Speaker Embedding Extractors Using Multi-Speaker Audio with Unknown Speaker Boundaries}
\name{Themos Stafylakis$^{1\dag}$, Ladislav Mo\v{s}ner$^{2\dag}$, Old\v{r}ich Plchot$^{2}$, Johan Rohdin$^{2,1}$, \\
Anna Silnova$^{2}$, Luk\'a\v{s} Burget$^2$, Jan ``Honza'' \v{C}ernock\'{y}$^2$
\thanks{\dag These authors contributed equally to this work.}}
\address{
$^1$Omilia - Conversational Intelligence, Athens, Greece\\
$^2$Brno University of Technology, Faculty of Information Technology, Speech@FIT, Czechia}
\email{tstafylakis@omilia.com, imosner@fit.vutbr.cz}

\begin{document}
  
\maketitle

\begin{abstract}
In this paper, we demonstrate a method for training speaker embedding extractors using weak annotation. More specifically, we are using the full VoxCeleb recordings and the name of the celebrities appearing on each video without knowledge of the time intervals the celebrities appear in the video. We show that by combining a baseline speaker diarization algorithm that requires no training or parameter tuning, a modified loss with aggregation over segments, and a two-stage training approach, we are able to train a competitive ResNet-based embedding extractor. Finally, we experiment with two different aggregation functions and analyze their behaviour in terms of their gradients.  
\end{abstract}

\section{Introduction}
Speaker embeddings are low-dimensional vectors that capture the voice characteristics of a person. Their role in speech technology is vital, especially in speaker recognition, diarization, separation, as well as in multi-speaker text-to-speech and voice conversion. Speaker embeddings are extracted with neural network architectures, such as TDNNs and ResNets \cite{snyder2018Xvectors,ecapa-tdnn,chung2018voxceleb2}. In a typical setting, the networks are trained in a supervised way, with a training set consisting of several thousands of speakers and multiple recordings per speaker.

For speaker recognition applications, where the training set should contain high intrinsic and extrinsic within-class variability, the most commonly used publicly available corpora are (a) the NIST-SRE repository (distributed by LDC) \cite{cts-superset} and (b) the VoxCeleb dataset \cite{nagrani2017voxceleb, chung2018voxceleb2}. The former has been collected and updated during the last two decades and is primarily composed of telephone conversations between two speakers in a controlled setup. The latter is composed of about 7 thousand celebrities and each such recording is an excerpt of an interview available on YouTube. Its creation was based on an innovative pipeline, combining querying YouTube for videos of each celebrity, a ResNet50-based face recognizer \cite{resnet} that leverages photos of each celebrity, a SyncNet architecture \cite{chung2016out} which verifies that the voice comes from the celebrity appearing at a given frame (active speaker verification), and some minor human verification \cite{nagrani2017voxceleb}. The decision thresholds were tuned so that the precision of the search is maximized at the expense of recall.

The emergence of self-supervision methods in deep learning has also been applied to training speaker embedding extractors \cite{stafylakis2019self,cai2021iterative,thienpondt2020idlab,slavivcek2021phonexia,mun2021bootstrap}. Several approaches have been examined, some of which employ an audiovisual setting \cite{cai2022incorporating,brown2022voxsrc}. 

It should be noted though that the whole pipeline of VoxCeleb is heavily dependent on the existence of audiovisual material, and therefore is not applicable to recordings that do not contain a video component (e.g. telephone conversations, radio broadcasts, etc.). Furthermore, the pipeline inevitably rejects potentially useful portions of the audio of the celebrity, coming from video excerpts during which the celebrity's face does not appear in the video.

In this work, we explore a novel way of training speaker embedding extractors on datasets containing conversational data (i.e. recordings with more than one speaker and a single channel). Contrary to self-supervised approaches, we assume that the speaker-of-interest labels are available on a recording level. What we consider as unknown are the speech chunks of recordings the celebrities appear in. We believe that this is a realistic setup since labels can be fairly easily obtained, e.g. using queries with celebrity names to a database, similarly to VoxCeleb.  

Our method consists of two stages. In the first one, the task is to estimate which chunks of a VoxCeleb2 full-length (uncut) recording belong to the target celebrity. In the second stage, the extractor is trained using these chunks with standard supervised training. The first stage, which is obviously the challenging one, involves a very basic diarization algorithm in order to estimate clusters. We assume that the clusters are of high speaker purity and low speaker coverage, i.e. speakers are typically clustered in more than one cluster. To find the chunks that correspond to the target celebrity, we train from scratch a speaker embedding extractor with Cross-Entropy,  Additive Angular Margin (AAM, \cite{arcface}) loss and a classification head over training celebrities. We create minibatches by randomly sampling segments from each cluster, and we feed those segments to the embedding network. The network extracts a 2D tensor containing the segment-level logits (i.e. number of training celebrities times clusters), which we aggregate across the cluster dimension using either max pooling or its soft version (the well-known log-sum-exp operator). This yields a vector of aggregated logits of size equal to the number of training celebrities which is fed to the softmax layer, and AAM loss is applied using the target celebrity as label. Once the model is trained, we keep those chunks that are classified by the network as belonging to the target celebrity. Finally, based on these chunks, we train a new extractor from scratch, showing competitive performance compared to one trained on the original VoxCeleb segments.

\section{Related work}
% In real-world applications, obtaining labels of plentiful data for strongly supervised training is a costly and resource-demanding process. On the other hand, it turns out to be much easier and cheaper to acquire data points with incomplete, inexact, or inaccurate labels, which has attracted the attention of the research community \cite{weak-sup}. Approaches to training classifiers and regressors utilizing the three aforementioned types of labels can be regarded as weakly supervised. In our study, we aim at employing inexact supervision, which generally deals with data having coarse-grained labels. The task at hand is commonly referred to as \emph{multi-instance learning} \cite{multi-instance} where labels are known for input \emph{bags}. Bags contain individual \emph{instances} (hence the name). Our instances are clustered audio segments. Bags represent whole files for which we have speaker labels. According to a multi-instance learning taxonomy \cite{multi-instance-tax}, the presented approach falls within an embedded space paradigm. It can be considered a vocabulary-based method with a set of speaker prototypes being vocabulary. Interestingly, even though a bag classification is enforced during training, it is the instance embedding that is utilized in test time.
In real-world applications, obtaining labels of plentiful data for strongly supervised training is a costly and resource-demanding process. It turns out to be much easier and cheaper to acquire data points with incomplete, inexact, or inaccurate labels \cite{weak-sup} leading to \emph{weakly supervised} training. In our study, we aim at employing \emph{inexact} supervision, which generally deals with data having coarse-grained labels. The task at hand is commonly referred to as \emph{multi-instance learning} \cite{multi-instance} where labels (speaker identities) are known for input \emph{bags} (recordings). Bags contain individual \emph{instances} (audio segments), hence the name. According to a multi-instance learning taxonomy \cite{multi-instance-tax}, the presented approach falls within vocabulary-based methods of an embedded space paradigm. Despite enforcing bag classification in training, instance embeddings are of interest in test time.

As summarized in \cite{multi-instance-tax}, multi-instance learning has been applied to tasks of multiple fields. In speech processing, weak supervision in the context of automatic speech recognition was explored in \cite{weaklyASR}. The authors trained a word-level acoustic model with the bag-of-words annotations. Their and our approaches are similar in terms of aggregation via log-sum-exp operation. The closest to our approach is that presented in \cite{Tanel2018, Tanelarxiv}, where the goal is speaker identification and verification. In contrast to our method, \cite{Tanel2018} requires a trained diarization model which provides audio segments for the i-vector \cite{iVector} extraction. %These are used as input to a speaker classifier trained to optimize an objective function taking into account weak labels. Eventually, the resulting classifier labels data for the final model training.
I-vectors are used as input to a speaker classifier trained with an objective encompassing weak labels. Eventually, the classifier labels data for the final model training. 

Finally, there are works and datasets that deal with the problem of multi-speaker enrolment or test recordings \cite{snyder2019speaker,mclaren2016speakers}. However, the extractor is trained on fully-supervised methods and their focus is on scoring rather than on training. 

\begin{figure}[t]
  \centering
  \includegraphics[width=\linewidth]{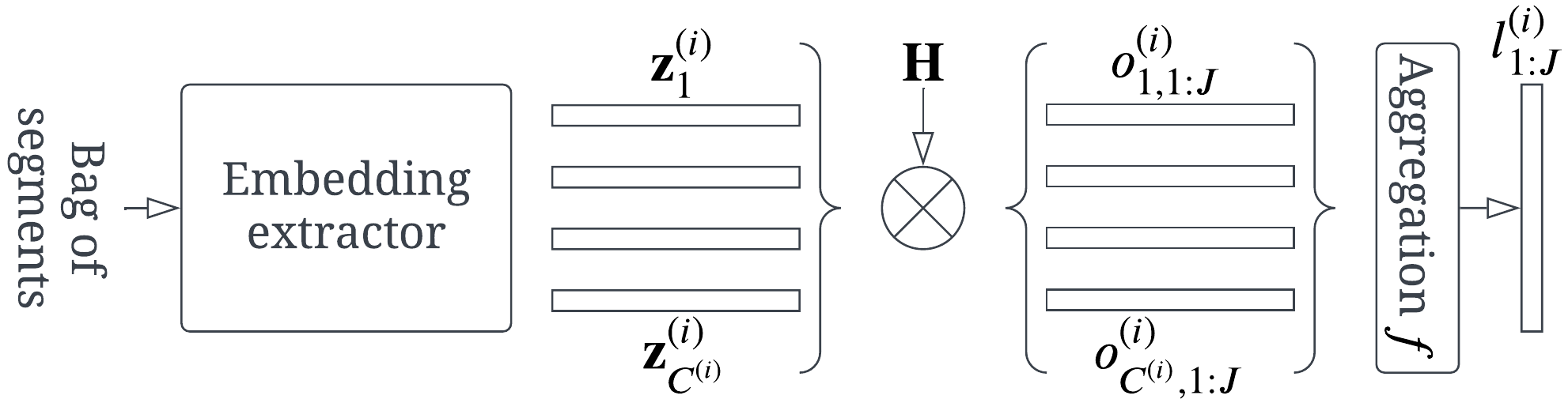}
  %\vspace{-2.5em}
  \vspace{-1.5em}
  \caption{Proposed method to weakly supervised training of a speaker embedding extractor. The $\otimes$ symbol stands for matrix multiplication.}
  \label{fig:architecture}
  \vspace{-1.5em}
\end{figure}

\section{Proposed method}
\subsection{Assumptions}
We assume that we have a set of $R$ recordings for training an embedding extractor. Moreover, we assume that we know all the celebrities appearing in each recording\footnote{We removed recordings with more than one celebrity labels, although our method can be fairly easily extended to include such cases.}. We do not know though where these celebrities appear in the recording, for how long they speak, or how many other speakers participate in the recording. We denote the set of celebrities by ${\cal J}$ and $J=|{\cal J}|$ is their number. Note that we use the term \emph{celebrities} as a reference to VoxCeleb, to describe those speakers included in the training set and the classification head of the network.    
\subsection{Baseline Speaker Diarization}
There are several speaker diarization algorithms that can yield state-of-the-art performance, ranging from methods that combine neural approaches with probabilistic clustering (e.g. \cite{DiezInter19,Silnova2020}) to fully end-to-end neural approaches (e.g. \cite{horiguchi2020end}). A weakness of these methods is their dependence on annotated datasets in order to train. As the purpose of this paper is to minimize the need for annotation, we will employ a very baseline diarization approach, namely (a) speaker change detection, followed by (b) Bayesian Information Criterion (BIC) based agglomerative hierarchical clustering with Gaussians on MFCC features for modeling segments, followed by (c) Viterbi-based boundary refinement using (maximum-likelihood trained) Gaussian Mixture Models \cite{tranter2006overview}. No pretrained models (such as Universal Background Model) are used for diarization. The SIDEKIT software is used for diarizing the recordings \cite{broux2018s4d}. We use the default hyperparameters, which tend to overestimate the number of speakers in a recording. This is needed for our algorithm in order to avoid merging chunks of different speakers in the same cluster, and to ensure that during training, at least one sampled segment corresponds to the target celebrity of the recording (see Section \ref{sec:aggregation}). The clustering is fixed during training, i.e. we do not re-cluster the chunks while training the extractor, as some self-supervised algorithms do \cite{cai2021iterative,thienpondt2020idlab}.

\subsection{Training the extractor with score-level aggregation functions}
\subsubsection{Creating minibatches}

Assume that the $i$th recording has been split into the set ${\cal S}^{(i)}$ of \emph{chunks} (excerpts). Division of ${\cal S}^{(i)}$ to disjoint sets ${\cal S}^{(i)}_c$ is determined by diarization, where $c \in {\cal C}^{(i)}$ and ${\cal C}^{(i)}$ is an index set of clusters. $|{\cal C}^{(i)}|= C^{(i)}$ varies with $i$.
The $i$th recording is represented in a minibatch by $\left\{g(S^{(i)}_c)\right\}_{c \in {\cal C}^{(i)}}$, where $S^{(i)}_c \in {\cal S}^{(i)}_c$ and $g(\cdot)$ cuts out a fixed-length \emph{segment}. Note that every batch can comprise a different number of recordings as a result of a different number of clusters.

\subsubsection{Aggregating over the cluster dimension}
\label{sec:aggregation}
In the forward pass, each segment is processed individually (i.e. independently of the recording it belongs to) including the statistics pooling layer, the embedding, and the dot-products of the embeddings with the linear layer of the classification head $\mathbf{H}=[\h_1,\h_2,\ldots,\h_{J}]$, where $\h_{j}$ denotes the set of weights for speaker $j$. This operation yields a 2D tensor (ignoring the batch dimension) of similarities between segments and celebrities. Tying between segments of the same recordings is implemented via an aggregation operation, which transforms the tensor into a vector of size $J$, indicating the similarity between the recording and celebrities. 

The segment-level similarities are defined as $\o_j^{(i)} = \{o_{c,j}^{(i)}\}_{c \in {\cal C}^{(i)}}$, where $o_{c,j}^{(i)}=\h_j^\intercal \z^{(i)}_{c}$ and $\z^{(i)}_{c}$ is the embedding of the example belonging to the $c$th cluster. The recording-level similarities $\{l_{j}^{(i)}\}_{j \in {\cal J}}$ are derived by an aggregation function:
\begin{equation}
l_{j}^{(i)} = f(\o_{j}^{(i)};\tau)
\end{equation}
as shown in Fig. \ref{fig:architecture}. We refer to these recording-level similarities as logits, since they are the ones passed to the SoftMax function
\begin{equation}
\label{eq:logits}
p\left(j|{\cal S}^{(i)}\right)= \frac{\exp(sl^{(i)}_{j})}{\sum_{j' \in {\cal J}}\exp(sl^{(i)}_{j'})}, 
\end{equation}
where $s$ is the scale of the AAM loss, while the target speaker logit is penalized by the margin of the AAM loss \cite{arcface}.
\subsubsection{Aggregation functions}
The most commonly used aggregation functions are the average and max pooling. However, both functions have certain undesired properties for our task. Average pooling would encourage the embeddings of all segments of the $i$th recording to be close to the representation $\h_j$ of the target celebrity of the recording. This property makes it inadequate for the task since only a fraction of the segments belongs to the target celebrity.

Max pooling seems to be more adequate since it focuses only on the segment having the highest similarity with the target celebrity. However, max pooling propagates gradients only through this highest-similarity segment, ignoring all other segments of the recordings. Furthermore, max pooling might make the network hard to train in the warm-up phase, when the network is randomly initialized.

Given the above, we also experiment with a soft version of max pooling, the log-sum-exp:

\begin{equation}
\label{eq:LSE}
l^{(i)}_{j} = f_{\mbox{LSE}}\left(\o_j^{(i)};\tau \right) = \tau \log \frac{1}{C^{(i)}}\sum_{c \in {\cal C}^{(i)}} \exp{\left(\frac{1}{\tau} o_{c,j}^{(i)}\right)}, 
\end{equation}
where $\tau > 0$ is a temperature. Note that for $\tau\xrightarrow{} 0^{+}$ we obtain max pooling, i.e. $f_{\mbox{LSE}}\left(\o_j^{(i)};\tau \right) \xrightarrow{} \max_{c}\left(\{o_{c,j}^{(i)}\}_{c \in {\cal C}^{(i)}}\right)$.

\subsubsection{Gradients and loss function}
It is interesting to derive the gradients when log-sum-exp is applied. Let $j^{*}$ be the (ground-truth) celebrity of the $i$th recording and let $L^{(i)}$ be the CE loss.  
By applying the chain rule, we get
\begin{equation}
\frac{\partial L^{(i)}}{\partial o_{c,j}^{(i)}} = p(c|j,{\cal S}^{(i)})\left[p(j|{\cal S}^{(i)}) - \delta_{j,j^{*}}\right], 
\end{equation}
where $\delta_{j,j^{*}}$ is the Kronecker delta and
\begin{equation}
p(c|j,{\cal S}^{(i)}) = \frac{ \exp{\left(\frac{1}{\tau} o_{c,j}^{(i)}  \right)} }{ \sum_{c' \in {\cal C}^{(i)}}  \exp{\left(\frac{1}{\tau} o_{c',j}^{(i)}\right)} }.
\end{equation}
To derive the above, we used the fact that the gradient of $l_j^{(i)}$ w.r.t. $o_{c,j}^{(i)}$ is $p(c|j,{\cal S}^{(i)})$. As we observe, the error signal $\epsilon^{(i)}_{j,j*} = p(j|{\cal S}^{(i)}) - \delta_{j,j^{*}}$ is distributed between the clusters proportionally to $p(c|j,{\cal S}^{(i)})$. Moreover, for utterances not containing the $j$th speaker, the error is proportional to $p(j|{\cal S}^{(i)})$, meaning that the norm of the gradient is small for recordings not containing speakers similar to the $j$th celebrity. 

Note also that if we consider length-normalized speaker representations and embeddings, i.e. $\norm{\h_j} = \norm{\z} = 1$, then $-1 \leq f\left(\o_j^{(i)};\tau \right) \leq +1$ for both aggregation functions we examine, making AAM loss applicable \cite{arcface}.

\begin{table}[tb]
    \centering
    \caption{Comparison of VoxCeleb2 derivatives.}
    \vspace{-0.5em}
    \begin{tabular}{l c r r}
        \toprule
        \tb{Dev. set} & \textbf{Celebrities} & \textbf{Recordings} & \textbf{Hours}\\
        \midrule
        \tb{Original} & 5,994 & 145,569 & 2,369.0 \\
        \tb{Restricted} & 5,987 & 110,940 & 1,884.3 \\
        \tb{Uncut} & 5,987 & 110,940 & 10,211.6 \\
        \tb{Self-labeled} & 5,935 & 88,084 & 4,483.8 \\
        \bottomrule
    \end{tabular}
    \label{tab:vox_comp}
    \vspace{-1.5em}
\end{table}

\subsection{Selecting chunks from the celebrities}
\label{sec:segment-selection}

Once training of the extractor is completed, we need to select the chunks corresponding to the target speaker. There are several methods that can be employed for this task. We implement a simple selection policy, where each chunk (as defined by the segmentation and boundary refinement algorithms of speaker diarization without the hierarchical clustering) is considered as coming from the celebrity of the recording if and only if the network classifies it as such. That is,
\begin{equation}
\label{eq:segment-selection}
S^{(i)} \in \hat{{\cal S}}_{j^*} \,\,\,\, \mbox{iff} \,\,\, \mbox{argmax}_j\left( \{l_j^{S^{(i)}}\}_{j\in {\cal J}}\right)  = j^*,
\end{equation}
where $S^{(i)}$ is a chunk from the $i$th recording, $\hat{{\cal S}}_j$ is the set of estimated chunks for the $j$th celebrity, and $j^*$ is the celebrity of the $i$th recording. The chunk selection stage is performed after training, so neither an aggregation function nor the AAM margin are applied. We simply classify all segments based on the logits denoted by $\{l_j^{S^{(i)}}\}_{j\in {\cal J}}$.   

\begin{table*}[tb]
    \centering
    \caption{Speaker verification results with models trained in strongly and weakly supervised fashion. $m$ stands for the AAM margin, $\uparrow$ means scheduling up to the value to the right of the arrow. LSE means log-sum-exp. $P_{tar} = 0.05$ for MinDCF. EER is in \%.}
    \vspace{-0.5em}
    \resizebox{\linewidth}{!}{%
    \begin{tabular}{l l l l l c c c c c c c}
        \toprule
        & \multirow{2}{*}{\tb{Supervision}} & \multirow{2}{*}{\tb{Data}} & \multirow{2}{*}{\tb{$m$}} & \multirow{2}{*}{\tb{Aggregation / $\tau$}} & \multirow{2}{*}{\tb{\shortstack{Speaker\\ repres.}}} &
        \multicolumn{2}{c}{\tb{VoxCeleb1-O}} & \multicolumn{2}{c}{\tb{VoxCeleb1-E}} & \multicolumn{2}{c}{\tb{VoxCeleb1-H}} \\
        & & & & & & \tb{EER} & \tb{MinDCF} & \tb{EER} & \tb{MinDCF} & \tb{EER} & \tb{MinDCF} \\
        \midrule
        s1 & strong & restricted & $0.1$ & -- & 1 & 1.56 & 0.111 & 1.56 & 0.100 & 2.81 & 0.164 \\
        s2 & strong & restricted & $\uparrow 0.3$ & -- & 1 & 1.24 & 0.074 & 1.34 & 0.087 & 2.46 & 0.142 \\
        \midrule
        s3 & weak & uncut & $0.1$ & max pooling / -- & 1 & 4.75 & 0.283 & 5.05 & 0.300 & 7.44 & 0.395 \\
        s4 & weak & uncut & $0.1$ & LSE / $0.5$ & 1 & 6.03 & 0.381 & 6.39 & 0.391 & 9.99 & 0.529 \\
        s5 & weak & uncut & $0.1$ & LSE / $0.5 \rightarrow 0.1$ & 1 & 4.53 & 0.286 & 4.86 & 0.292 & 7.31 & 0.401 \\
        \midrule
        s6 & pseudo & self-labeled & $\uparrow 0.3$ & -- & 1 & 1.73 & 0.114 & 1.80 & 0.114 & 3.15 & 0.184 \\
        s7 & pseudo & self-labeled & $\uparrow 0.3$ & -- & 2 & 1.58 & 0.103 & 1.82 & 0.118 & 3.23 & 0.184 \\
        \bottomrule
    \end{tabular}
    }
    \label{tab:results}
    \vspace{-1.5em}
\end{table*}
        % strong & restricted & -- & 1 & 1.24 & 0.074 & 1.34 & 0.087 & 1.46 & 0.086 & 1.49 & 0.095\\
        % \midrule
        % weak & uncut & max & 1 & 4.75 & 0.283 & 5.05 & 0.300 & 4.79 & 0.289 & 5.03 & 0.302 \\
        % weak & uncut & LSE / $0.5$ & 1 & 6.03 & 0.381 & 6.39 & 0.391 & 5.73 & 0.363 & 6.03 & 0.368 \\
        % weak & uncut & LSE / $0.5 \rightarrow 0.1$ & 1 & 4.53 & 0.286 & 4.86 & 0.292 & 4.29 & 0.269 & 4.75 &  0.282 \\

\section{Experimental setup}
\subsection{Speaker embedding extractor and classifier}
All presented models trained in a weakly supervised fashion share the same structure: per-segment embedding extractor, an aggregation function with similarity computation, and AAM-based logit computation \cite{arcface}.

The backbone of embedding extractors is based on the ResNet34 architecture \cite{resnet}, widely used in the field of speaker verification \cite{brown2022voxsrc}. 400 frames of 80-dimensional log Mel-filter bank energy (fbank) features constitute the input to the model. Individual stages of the network comprise 2D convolutional layers with 64, 128, 256, and 256 filters each. The building blocks of stages follow the pre-activation structure \cite{full-preact}. Instead of a standard batch normalization, we opt for instance normalization \cite{ulyanov-inst-norm}. The rationale behind this choice is as follows: During training, segments do not need to contain only speech due to a simple diarization. Statistics collected by batch normalization layers are in turn also affected by non-speech content, which is in contrast with the test phase. The resulting embeddings are 256-dimensional. The embedding extraction is followed by aggregation described in Section \ref{sec:aggregation}.

Thanks to the favorable property of the aggregation functions -- values bounded in the interval [-1, 1] -- the output of the preceding stage can be considered as similarity scores (analogous to the cosine similarity computed in fully supervised training). Therefore, logits for the CE loss function can be computed according to margin-based methods. In our case, the margin is introduced following the AAM formulation. In all of our experiments, we use a scale $s=30$, see Eq. (\ref{eq:logits}). It is a common practice to increase the margin during training. Since we experimentally found it challenging to schedule it in weakly supervised settings, we keep it fixed at 0.1 (unless explicitly stated).

The models are trained using SGD with a momentum of 0.9. To prevent from diverging in the initial stage of such challenging training, we employ a learning rate warm-up. After reaching a value of 0.2, the learning rate is scheduled according to the loss on the cross-validation set comprising 256 speakers whose recordings were extracted from the training set.

\subsection{Strongly and weakly labeled datasets}
For the sake of comparison, we choose an exemplary training corpus termed \emph{VoxCeleb2 Uncut dev}. It has been acquired by downloading a complete audio material of recordings present in VoxCeleb2 dev. We stress that we use only full-length audio recordings without a video content of the corresponding clips. It might seem wasteful not to use another available modality, but we target a more general goal -- utilization of weakly labeled audio-only data, which is much easier to obtain compared to strongly labeled segments.

Since some video clips are no longer available, \emph{VoxCeleb2 Uncut} is restricted compared to the original VoxCeleb2 in terms of speakers and recordings. The comparison is presented in Table \ref{tab:vox_comp}. To provide a fair comparison with baselines trained in a standard supervised way, we limit the speakers and recordings of VoxCeleb2 to match the \emph{uncut} version -- \emph{VoxCeleb2 Restricted dev}. In both weakly- and strongly-supervised training, the data is sampled at 16~kHz. Online reverberation and Musan noise \cite{musan} augmentations are applied to boost robustness.

\section{Experiments}
The evaluation is performed on the standard sets of verification trials -- VoxCeleb1-O, E, and H. We present results in terms of equal error rate (EER [\%]) and minimum detection cost (MinDCF), with prior probability of target trials $P_{tar}=0.05$.

\subsection{Strong vs. weak supervision}
In the first section of Table \ref{tab:results}, we present the results of a standard supervised training with both scheduled (0.1 $\rightarrow$ 0.3) and constant (0.1) margin $m$ of AAM. We show how enforced between-speaker separability can improve performance to stress the potential of proposed weakly supervised training, which utilizes a constant margin of 0.1. As per our experimentation, however, more sophisticated treatment of margin within our framework is required. We note that the models' performance is affected by a limited size of \emph{VoxCeleb2 Restricted dev} used for training.
% In the first section of Table \ref{tab:results}, we present the results of a standard supervised training. During training, the AAM margin is scheduled to increase from 0.1 to 0.3 (as opposed to weakly trained systems). We note that the model performance is affected by a restricted size VoxCeleb2 dev used for training.

The next set of rows of Table \ref{tab:results} clearly shows that proposed weakly supervised training is viable and already leads to reasonably well-performing models. Obviously, multi-instance learning is much harder than training with strong supervision, which is reflected by a deterioration of the results compared to the baseline. We compare two possible aggregation functions: max pooling and log-sum-exp. Despite its limitations described in Section \ref{sec:aggregation}, the max pooling is a plausible option and even outperforms log-sum-exp with a constant temperature of 0.5. We assume the worse performance yielded by $\tau=0.5$ is due to over-smoothness. On the other hand, scheduled $\tau$ changing from 0.5 to 0.1 eases training, and the aggregation gets closer to the maximum in later stages of training. We observe that a soft version of the maximum aggregation can lead to improved performance. As $\tau$ has a significant impact, it calls for further exploration. We suspect that better scheduling has the potential to boost performance.

\subsection{Training with self-labeled data}
Having trained models in a weakly supervised fashion, we use the best one (s5) to select chunks of the \emph{uncut} set according to~(\ref{eq:segment-selection}). The procedure from Section \ref{sec:segment-selection} provides the \emph{self-labeled} training set with parameters presented in Table \ref{tab:vox_comp}. Even though some speakers and recordings have been lost, it is 2.4 times larger than the \emph{restricted} set. We proceed with supervised training with obtained pseudo labels where aggregation function is no longer required.

Table \ref{tab:results} shows that the resulting system s6 greatly outperforms the model used for labeling. At the same time, its performance gets close to that of strongly supervised models.

As a result of inaccurate diarization and the chunk selection approach, the \emph{self-labeled} set contains non-speech signals or chunks with speakers without labels (e.g. interviewers). To deal with label noise, we introduce two representations per speaker in the classification head $\mathbf{H}$ \cite{Deng2020sub-arcface,thienpondt2020idlab}. This allows it to learn true speaker and interference representations for each celebrity. Although a considerable improvement can be observed on VoxCeleb1-O (s6 vs s7), the results on the other two sets do not show a similar trend.

\section{Conclusions}
In this paper, we presented an approach to weakly supervised training of speaker embedding extractors with mere recording-level labels. We demonstrated our method using the full-length audio recordings of VoxCeleb2 clips. We initially trained an extractor equipped with an aggregation function to estimate chunks belonging to the target celebrity. 
Using these chunks, we trained a model from scratch, attaining performance close to the model trained on the original VoxCeleb segments.

Encouraged by the results, we plan to extend the approach by introducing alternative ways of aggregation (such as attention) and leverage recordings containing more than one celebrity. We will also explore combinations of weak and self-supervision for training the model. Finally, we will experiment with applying a second, embedding-based diarization stage (using embeddings from the first extractor) and make use of other (non-celebrity) speakers in training.

\section{Acknowledgements}
The work was supported by Czech Ministry of Interior project No. VJ01010108 "ROZKAZ", Czech National Science Foundation (GACR) project NEUREM3 No. 19-26934X, European Union’s Horizon 2020 project No. 833635 ROXANNE and Horizon 2020 Marie Sklodowska-Curie grant ESPERANTO, No. 101007666. Computing on IT4I supercomputer was supported by the Czech Ministry of Education, Youth and Sports from the Large Infrastructures for Research, Experimental Development and Innovations project "e-Infrastructure CZ – LM2018140". 

%{\color{red}The authors do not see any significant ethical or privacy concerns that would prevent the processing of the data used in the study. The datasets do contain personal data, and these are processed in compliance with the GDPR and national law. ???}

\bibliographystyle{IEEEtran}
\bibliography{biblio}

\end{document}